\documentclass[aps,english,superscriptaddress,preprintnumbers,reprint,footinbib,amsmath,amssymb,prl]{revtex4-2}
\usepackage[version=3]{mhchem}
\usepackage{graphicx}
\usepackage{dcolumn}
\usepackage{bm}
\usepackage{hyperref}
\hypersetup{
    colorlinks=true,
    linkcolor=blue,
    filecolor=blue,
    urlcolor=blue,
   citecolor=blue,
}
\usepackage{enumitem}

\begin{document}


\title{Bound photonic pairs in 2D waveguide quantum electrodynamics}

\author{Y. Marques}
\affiliation{ITMO University, St.~Petersburg 197101, Russia}

\author{I. A. Shelykh}
\affiliation{ITMO University, St.~Petersburg 197101, Russia}
\affiliation{Science Institute, University of Iceland, Dunhagi-3, IS-107,
Reykjavik, Iceland}

\author{I. V. Iorsh}
\email[corresponding author:]{iorsh86@yandex.ru}
\affiliation{ITMO University, St.~Petersburg 197101, Russia}

\date{\today}

\begin{abstract} 
We theoretically predict the formation of two-photon bound states in a two-dimensional waveguide network hosting a lattice of two-level atoms. The properties of these bound pairs and the exclusive domains of the parameter space where they emerge due to the interplay between the on-site photon blockade and peculiar shape of polariton dispersion resulting from the long-range radiative couplings between the qubits are investigated in detail. In addition, we analyze the effect of the finite-size system on localization characteristics of these excitations.
\end{abstract}

\maketitle


\textit{Introduction.} The recent development of nanotechnology resulted in the appearance of unprecedented platforms for many-body quantum electrodynamics, consisting of quantum emitters coupled to propagating photons in waveguides~\cite{Roy2017,KimbleRMP2018,Tursch2019,sheremet2021Arxiv}.  Particular realizations of such waveguide quantum electrodynamics (WQED) systems, include structures based on artificial arrays of cold atoms~\cite{Corzo2019, Goban2015}, superconducting qubits~\cite{vanLoo2013,Mirhosseini2019}, quantum dots~\cite{Foster2019} and solid-state vacancies defects~\cite{Sipahigil2016}. The exquisiteness of WQED systems is that they demonstrate an interplay of strong light-matter interaction, chirality, and long-range radiative couplings between quantum emitters arising from the exchange of the propagating photons. The combination of these features gives rise to a plethora of fascinating physical phenomena, including collective super-radiance and sub-radiance ~\cite{Ke2019,kornovan2019extremely,Albrecht2019a,Henriet2019,Zhang2019,Ke2019,Zhang2019b,Fatih2019,Fatih2020}, emergence of unconventional topological phases~\cite{kim2020quantum,Chang2020}, quantum chaos~\cite{Poshakinskiy2021}, and promotes insightful developments for emergent quantum technologies.

Long-range coherent photonic propagation in a waveguide  couples all emitters together and leads to the formation of collective polaritonic excitations~\cite{sheremet2021waveguide}. Since a given emitter can be excited only by a single photon,  such structure represents an example of a strongly correlated system ~\cite{Birnbaum2005}. One of its most compelling properties is the possibility of the formation of unconventional multi-photon bound states, attracting the growing interest of theoretical researchers~\cite{Poddubny2020,Zhang2020,PhysRevX.10.031011,Zhong2020,Zhong2021}.  However, up to date, most of the efforts were dedicated to the consideration of one-dimensional (1D) set-ups, since they were the only ones accessible experimentally. However, very recently a two-dimensional (2D) array of waveguide-coupled array of transmon qubits was realized~\cite{gong2021quantum}, which makes actual the task of the proper theoretical description of strongly correlated WQED in higher dimensions. In this context, the fundamental question appears: whether bound two-polariton states exist in 2D at all, and if they do, what are their localization characteristics. 

\begin{figure}[t]
	\centerline{\includegraphics[width=3.0in,keepaspectratio]{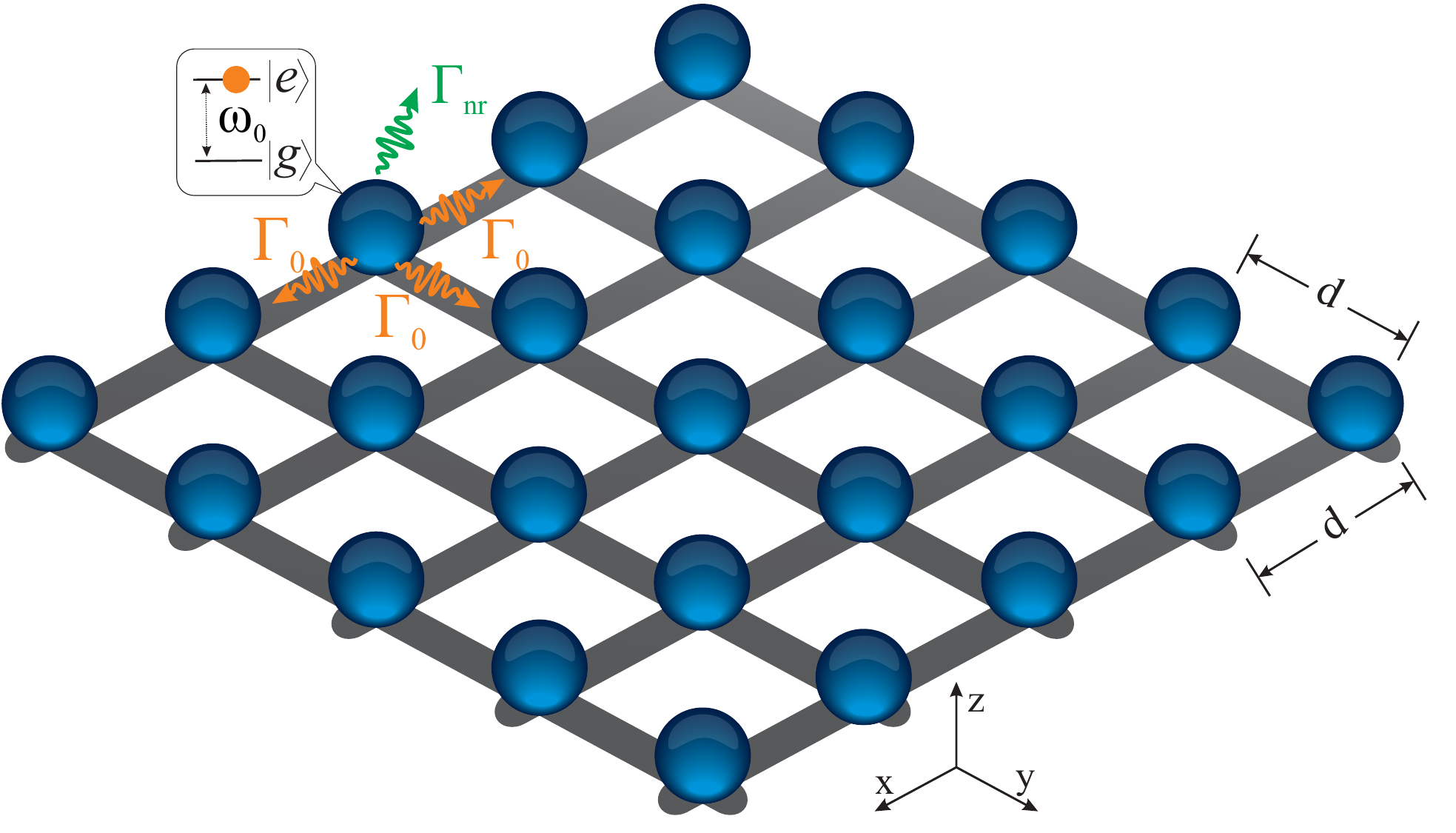}}
	\caption{\label{fig:Fig1} Sketch of the considered setup consisting of a square lattice of regularly spaced qubits placed over a two-dimensional network of waveguides. The qubits, two-level atoms with resonant frequency $\omega_0$ between ground $\left|g\right\rangle$ and excited $\left|e\right\rangle$ states, couple with identical waveguides and display equivalent emission decay rate $\Gamma_0$ in both the $x$ and $y$ directions. The non-radiative emission decay  rate $\Gamma_{\textrm{nr}}$ addresses  the losses from the scattered photons outward the system.} \label{fig:1}
\end{figure}

In this Letter, we explore the formation of the bound two-polariton states in a 2D WQED set-up shown in Fig.~\ref{fig:1}. We show that the bound states indeed exist inside the band gap for the scattering states and establish their spatial profiles. We also demonstrate the characteristics of these polariton pairs in finite-size systems that can be detected in scattering experiments. 


\textit{Two-particle Hamiltonian.}  We consider the system schematically shown in Fig.~\ref{fig:1}. It consists of an $N \times N$ square lattice of qubits located at the nodes of a network composed of a set of horizontal and vertical identical one-dimensional waveguides in the $xy$ plane. Each qubit, described as a dipole with resonant frequency $\omega_0$ between ground $\left|g\right\rangle$ and excited $\left|e\right\rangle$ states, couples with a pair of waveguides that support propagating light modes with linear dispersion with velocity $v$. Waveguide photonic modes can be integrated out in the Markovian regime~\cite{Albrecht2019a, Zhong2020, Ke2019}. One additionally suggests an equal probability of a qubit decay into each of the two waveguides, the corresponding 2D Hamiltonian is given by:
\begin{equation}
\mathcal{H}_{\textrm{eff}}^{\textrm{2D}} = \mathcal{H}_{\textrm{eff}}^{\textrm{1D}}\otimes I+I\otimes\mathcal{H}_{\textrm{eff}}^{\textrm{1D}},
\end{equation}
where $\mathcal{H}_{\textrm{eff}}^{\textrm{1D}}$ is the effective 1D Hamiltonian that describes each single waveguide array.  For each waveguide, the Hamiltonian describes an open quantum system, where the coherent exchange of photons enables an infinite-range qubit-qubit interaction, 
\begin{eqnarray}
\mathcal{H}_{\textrm{eff}}^{\textrm{1D}} & = & \sum_{m,n=1}^{N}H_{mn}b_{m}^{\dagger}b_{n} + \frac{\chi}{2} \sum_{m=1}^{N} b_{m}^{\dagger} b_{m}^{\dagger}b_{m}b_{m}\label{eq:H_1D} 
\end{eqnarray}
with $H_{mn}=(\omega_{0}-i\Gamma_{\textrm{nr}})\delta_{mn}-i\Gamma_{0}e^{i\varphi|m-n|}$, where $\Gamma_{0}$ and $\Gamma_{\textrm{nr}}$ respectively stand for the radiative and non-radiative decay rate of a single qubit, $\varphi=q_0 d$ represents the phase acquired by excitations with wave vector $q_0=\omega_{0}/v$
when traveling between two qubits spaced by $d$, the annihilation operators
$b_{m}$ account for the bosonized excitations of the qubits,  and $\chi$ stems from the effective on-site photon-photon repulsion. 

The Hamiltonian (\ref{eq:H_1D}) effectively describes the coherent and dissipative collective interaction of the guided modes through $i\Gamma_{0}e^{i\varphi|m-n|}$ and also the inherent losses stemming from photon emission to the free space which is modulated by the non-radiative decay rate $\Gamma_{\textrm{nr}}$. In particular, the waveguide supports guided modes that hardly decay into free space ($\Gamma_{\textrm{nr}}/\Gamma_{0} \ll 1$) for small array periods $d<\lambda_{0}/2$ ($\varphi < \pi$), where $\lambda_{0}= 2\pi c/\omega_{0}$ is the atomic wavelength. This atom-waveguide coupling efficiency is close to $99\%$ in systems with real atoms coupled to a fiber waveguide~\cite{Corzo2019} and even exceeds $99.9\%$ for superconducting qubits~\cite{Mirhosseini2019}. Even still, the qubits may present a strong dipole-dipole interaction in this same domain where the distance between them is less than the atomic wavelength.  Nonetheless, the dipole-dipole interaction is rapidly suppressed in fiber waveguides due to a concept known as selective radiance, where the frequency of the photons is tuned to enhance emission rate into the waveguide while suppressing the emission to vacuum ~\cite{sheremet2021waveguide, Asenjo2017}. Hence, the energy scale of dipole-dipole interaction is much smaller than the energy scale of the system $\Gamma_{0}$, and then can be safely disregarded.

As single two-level atoms are prevented to be excited by two identical photons at the same time due to the Pauli exclusion principle, the system lies on the so-called hard-core limit ($\chi\rightarrow\infty$)~\cite{Zhong2020, Ke2019, Poshakinskiy2021}, where the occupation of each qubit, restricted to either $0$ or $1$, leads to a picture where the light-matter excitations (polaritons) effectively exhibit fermionic behavior~\cite{Chang2008}.

To analyze the nature of two-particle excitations of the 2D lattice we need to solve the corresponding linear eigenvalue problem written as (see Supplementary Material for the details):
\begin{align}
2\varepsilon\psi_{ij,mn}=&H_{il}\psi_{lj,mn}+H_{jl}\psi_{il,mn}+H_{ml}\psi_{ij,ln} \nonumber\\
+H_{nl}\psi_{ij,ml}-&2\delta_{jn}H_{il}\psi_{lj,in}-2\delta_{im}H_{jl}\psi_{il,mj}, \label{eq:TP_2D}
\end{align}
where $\psi_{ij,mn}$ denotes the probability amplitude associated with the polariton pair, in which $i,j(m,n)$ indicates the position of first (second) polariton. The indices $i,m$ correspond to the $x$-coordinates positions, $j,n$ to $y$-coordinates.

For infinite periodic lattice, the polariton pair is characterized by the center of mass wave vector $\boldsymbol{K}=K_{x}\hat{e}_{i}+K_{y}\hat{e}_{j}$, so that  two-particle amplitudes can be written as:
\begin{eqnarray}
\psi_{ij,mn}&=&e^{iK_{x}(i+m)/2}e^{iK_{y}(j+n)/2}\Phi_{i-m,j-n},\label{eq:TP_K} 
\end{eqnarray}
with wave function of the relative motion $\Phi_{0,0}=0$ and $\Phi_{i-m,j-n}=\Phi_{m-i,j-n}$. Substituting Eq.~(\ref{eq:TP_K}) into Eq.~(\ref{eq:TP_2D}) and introducing the relative distances $d_{x}=i-m$ and $d_{y}=j-n$, we find the system of equations characterizing the relative motion of a polariton pair, given by
\begin{eqnarray}
\varepsilon_{\boldsymbol{K}}\Phi_{d_{x},d_{y}}&=&\sum_{l=-\infty}^{\infty}\left(H_{l,d_{x}}\Phi_{l,d_{y}}+H_{l,dy}\Phi_{d_{x},l}\right),\label{eq:ES_k} 
\end{eqnarray}
where $H_{l,d\lambda}=-i\Gamma_{0}\cos\left(K_{\lambda}\frac{d_{\lambda}-l}{2}\right)e^{i\varphi|d_{\lambda}-l|}$ for $ \lambda =x,y$. Solutions of Eq.~(\ref{eq:ES_k}) describe both the scattering states corresponding to the continuous part of the  spectrum, and also, under specific conditions, the formation of bound pairs.

\begin{figure}[t]
	\centerline{\includegraphics[width=3.4in,keepaspectratio]{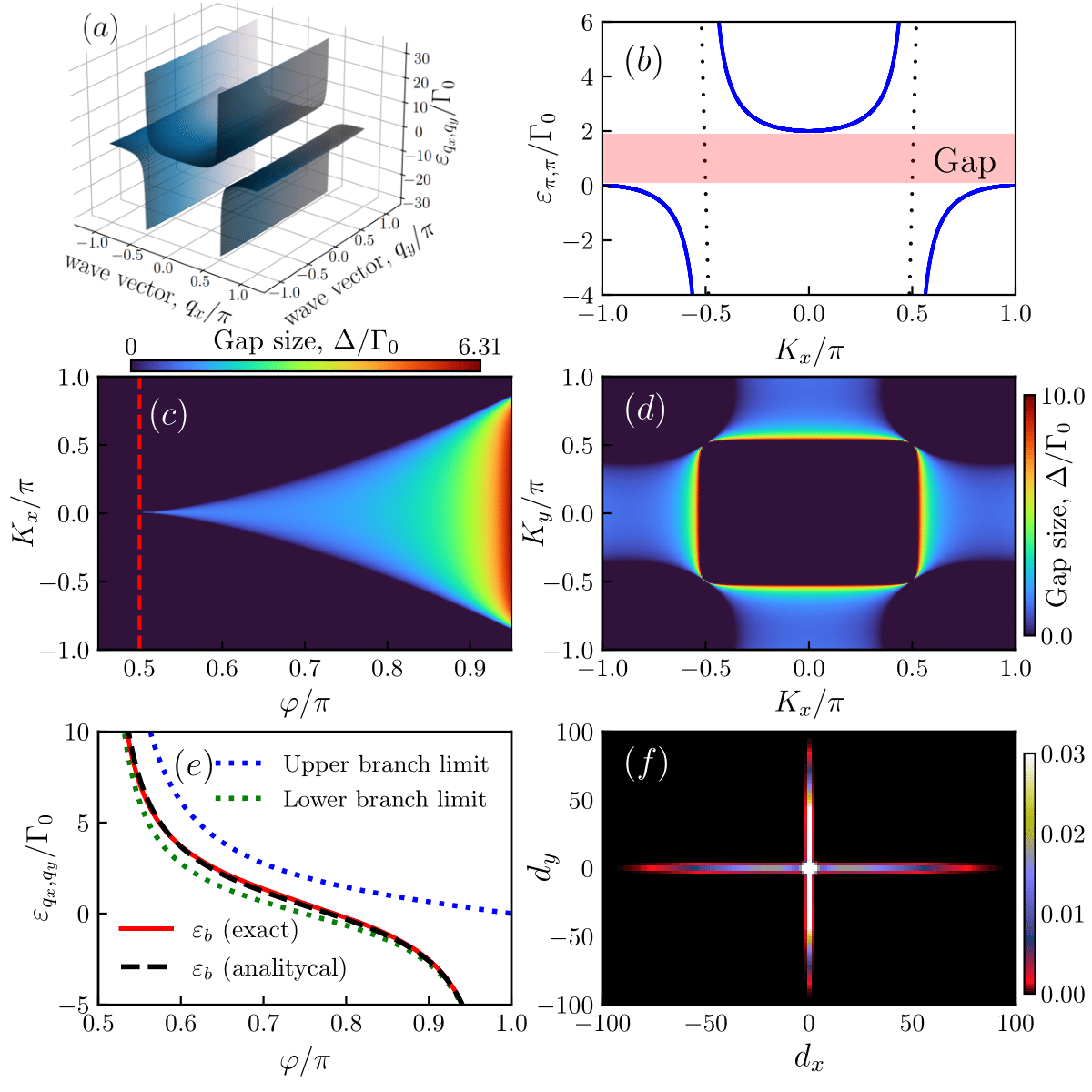}}
	\caption{\label{fig:2} (a) Polariton dispersion of the two-dimensional lattice for center of mass wave vectors $K_{x}=\pi$ and $K_{y}=0$. (b) 1D slice of the dispersion relation at $K_{y}=\pi$ considering isotropic polariton wave vectors $q_{x}=q_{y}=\pi$. The black dotted lines show the dispersion of light in pristine waveguides. (c), (d) The size of the energy gap in the polariton dispersion for $K_{x}=\pi$ and $\varphi=3\pi/4$, respectively. The vertical dashed line in panel (c) highlights the gap opening at $\varphi=\pi/2$. (e) Exact (solid red line) and analytical (black dashed line) bound state energy, where the blue and green dashed curves illustrates the lowest and highest energy values of the upper and lower polariton branches, respectively.  (f) Polariton spatial distribution $|\Phi_{d_{x},d_{y}}|^2$ (reescaled by 0.2). The polariton phase is set at $\varphi=3\pi/4$ in panels (a,c,e,f).}
\end{figure}

In order to obtain the scattering state dispersion relation, we move from the center of mass position basis to the relative motion wave vector ($-\pi<q_{x},q_{y}\leq\pi$) basis by performing a 2D cosine Fourier transform in Eq.(\ref{eq:ES_k}). As a result, the system dispersion relation equation is given by:
\begin{eqnarray}
2\varepsilon_{q_{x},q_{y}}&=&\Gamma_{0}(\frac{\sin\varphi}{\cos k_{1,x}-\cos\varphi}+\frac{\sin\varphi}{\cos k_{2,x}-\cos\varphi}) \nonumber\\&+&\Gamma_{0}(\frac{\sin\varphi}{\cos k_{1,y}-\cos\varphi}+\frac{\sin\varphi}{\cos k_{2,y}-\cos\varphi}).
\label{eq:DR} 
\end{eqnarray}
The total energy of a pair $2\varepsilon_{q_{x},q_{y}}$ is represented as a sum of the energies of non-interacting polaritons with wave vectors $k_{1(2),x(y)}=(q_{x(y)} \pm K_{x(y)})/2$. Its shape is determined by the phase $\varphi$ and the center of mass wave vector $\textbf{K}$ and it is shown in Fig.~\ref{fig:2}(a) for $\varphi=3\pi/4$, $K_{x}=\pi$, and $K_{y}=0$.

The impossibility of double occupation in a single qubit due to the on-site repulsion ($\chi\rightarrow\infty$) seems to suppress any possibility to observe bound state pairs. Nonetheless, the lattice has an infinite-range radiative coupling so that the polariton-polariton correlation, stemming from the on-site repulsion, is preserved all along the lattice. This is essential to the formation of bound states with repulsive interactions perceived by the negative effective mass regions in the dispersion relation shown in  Fig.~\ref{fig:2}(a) and Fig.~\ref{fig:2}(b). The different sign effective masses of polaritons at the center and the edge of the Brillouin zone allows for the formation of the in-gap bound two-polariton states with energies lying in the band gap even for the case of repulsive interactions. The creation of these finite-energy bound states by strong repulsive interaction has already been observed in Bose-Hubbard models in optical lattices~\cite{Winkler2006}.

As bound states arise as discrete in-gap states, the energy gap in the dispersion relation is the main characteristic that allows the formation of bound states. However, the existence of a gap is not guaranteed for any arbitrary values of $\varphi$ and $\textbf{K}$. Fig.~\ref{fig:2}(c) and Fig.~\ref{fig:2}(d) show the domains where two-polariton pair can be observed by revealing the gap size $\Delta$ in the dispersion relation. In Fig.~\ref{fig:2}(c), obtained for $K_{x}=\pi$, one can notice that bound states cannot be observed in the range of $\varphi \in [0;\pi/2]$, but arise for the parameters combination lying inside the cone-shaped domain. Fig.~\ref{fig:2}(d) maps $\Delta$ for values of the center of mass wave vector $\textbf{K}$ and fixed $\varphi=3\pi/4$.  The dispersion relation profiles where the energy gap is absent are shown in the Supplementary Materials. Given the complexity of Eq.~(\ref{eq:ES_k}), we fix the wave vectors $K_{x}=\pi$ and $K_{y}=0$ henceforth to achieve analytical expressions for the bound state energy and its corresponding wave functions.
\section{}
\textit{Two-polariton bound states.} To obtain the bound state energy $\varepsilon_{b}$, we assume that the condition for $\Phi_{0,0}=0$ is due to a scattering potential $\hat{v}=\varepsilon_{0}\left|\Phi_{0,0}\right\rangle \left\langle\Phi_{0,0}\right|$ with $\varepsilon_{0}\rightarrow+\infty$ applied to the unperturbed Hamiltonian characterizing nearly free polariton propagation with dispersion $\varepsilon_{q_{x},q_{y}}$. Within the Green's function formalism~\cite{economou2013}, the bound states correspond to the poles of the transfer matrix $\mathcal{T}=\hat{v}(\hat{I}-\hat{G}_{0}(d_x,d_y)\hat{v})^{-1}$. Hence, at the system origin, where the infinite scattering potential is present, the condition for bound states is given by $\hat{G}_{0}(0,0)=0$, i.e.,
\begin{eqnarray}
G_{0}(0,0)=\int_{-\pi}^{\pi}\int_{-\pi}^{\pi}\frac{dq_{x}dq_{y}}{\varepsilon_{b}-\varepsilon_{q_{x},q_{y}}}&=&0.\label{eq:GF_bare} 
\end{eqnarray}
Integral in Eq.~\eqref{eq:GF_bare} can be taken analytically, but results in a cumbersome expression involving elliptic integrals of the second and third kind. We therefore resort to the numerical solution. However, an approximate solution can be obtained within certain approximations. 

Namely, we first note that the bound state energy should lie in the band gap region, i.e., $\cot\varphi<\varepsilon_b<-\tan\varphi$ for $\varphi \in [\pi/2;\pi]$. Then, we can notice from Fig.~\ref{fig:2}(a) that dispersion along $q_y$ is weak. We thus can use the fact that $\varepsilon_{q_x,q_y}=\varepsilon(q_x)+\varepsilon'(q_y)$ and substitute average value of $\varepsilon'(q_y)$, $\langle \varepsilon'(q_y)\rangle=(2\pi)^{-1}\int dq_y \varepsilon'(q_y)$ in Eq.~\eqref{eq:GF_bare}. This would allow to obtain an approximate expression for the bound energy for $K_x=\pi,K_y=0$

\begin{eqnarray}
\varepsilon_{b} \approx 2\Gamma_{0}\cot(2\varphi)+\Gamma_{0}\mathrm{arctanh}(\cot(\varphi/2))\label{eq:BS_en} 
\end{eqnarray}
as it is shown by the solid line in Fig.~\ref{fig:2}(d). As can be seen, this approximation is very close to one given by numerical solution.

The bound polariton pair wave functions are obtained as
\begin{eqnarray}
\Phi_{d_{x},d_{y}}&=&\Gamma_{0} \int_{-\pi}^{\pi}\int_{-\pi}^{\pi}dq_{x}dq_{y}\frac{\cos\left(q_{x}d_{x}+q_{y}d_{y}\right)}{\varepsilon_{b}-\varepsilon_{q_{x},q_{y}}},\label{eq:WF} 
\end{eqnarray}
with discrete values of relative distances $d_{x},d_{y}=\{0,1,2,...\}$, except for $\Phi_{0,0}=0$. The solution details can be found in the Supplementary Materials.
The shape of the entire ensemble of solutions is presented Fig.~\ref{fig:2}(f) for arbitrary values of $d_{x}$ and $d_{y}$. We can see that the profile has cross-like structure with greater localization along $x$ axis ($K_x=\pi$) than along $y$ axis ($K_y=0$).
\begin{figure}[t]
	\centerline{\includegraphics[width=3.4in,keepaspectratio]{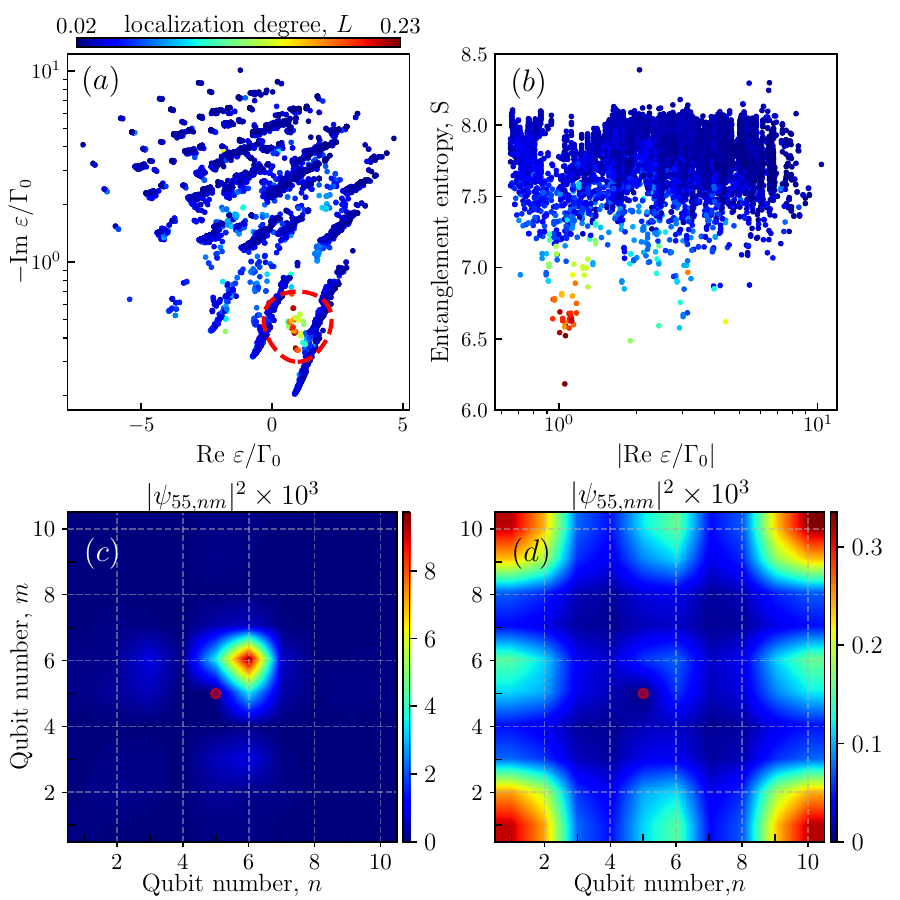}}
	\caption{\label{fig:3}  (a) The two-excitation eigenvalues of the finite structure system composed of $10 \times 10$ qubits. Each eigenvalue is characterized by the localization degree $L$ of its wave functions, in which $L \approx 1(0)$ stands for highly localized (delocalized) states. The dashed red circle highlights the highly correlated polariton states. (b) Entanglement entropy of two-polariton states. The polariton pairs spatial distribution $\left|\psi_{ij,nm}\right|^2$ of the highest localized state ($\varepsilon \approx 0.87-0.35i$) and the most subradiant state ($\varepsilon \approx 0.9-0.2i$),respectively. The polariton phase is set at $\varphi=3\pi/4$ and the non-radiative decay is fixed at $\Gamma_{\textrm{nr}}=0.1\Gamma_{0}$.}
\end{figure}
Generally, the eigenstates of WQED structures are classified by their collective decay ratio $\Gamma = -\textrm{Im}\varepsilon$ in comparison with the single qubit decay rate $\Gamma_0$, so that $\Gamma \sim N\Gamma_0$ correspond to superradiant states, $\Gamma \sim \Gamma_0$ correspond to bright states, and $\Gamma \ll \Gamma_0$ correspond to  subradiant states. Recently, new classes of eigenstates that emerge exclusively in multi-particle excitation regime, such as twilight~\cite{Yongguan2019}, chaotic~\cite{Poshakinskiy2021}, and also bound states~\cite{Poddubny2020,Zhang2020} were theoretically discovered. In contrast with infinite lattices where the polariton pairs are indeed bound states with an infinite lifetime in a qubit state, the finite lattice exhibits highly localized photon pairs with a finite lifetime which only become bound states when the periodic lattice limit is met. As experimental set-ups achieve a limited amount of qubits, it is relevant to explore finite systems and understand whether the highly correlated polariton pair are the most subradiant states of the system and what are the profiles of their spatial distribution.

This class of highly correlated polariton pairs is identified by its degree of localization $L$ based on the mode volume in optical cavities~\cite{Kristensen2014} and defined as 
\begin{eqnarray}
L&=& \frac{\sum_{m,n}\Psi_{m,n}^2}{ (\sum_{m,n}\Psi_{m,n})^2},\label{eq:S} 
\end{eqnarray}
with
\begin{eqnarray}
\Psi_{m,n}&=&\frac{1}{N^2}\sum_{i,j}(|\psi_{i,j,i+n,j+m}|^2 +|\psi_{i,j,i+n,j-m}|^2 \nonumber \\ &+& |\psi_{i,j,i-n,j+m}|^2 +|\psi_{i,j,i-n,j-m}|^2),\label{eq:S} 
\end{eqnarray}
where the states $\psi_{i,j,n,m}$ and the system eigenvalues $\varepsilon$ are obtained by direct diagonalization of Eq.~\ref{eq:TP_2D}. For the case of a state where both excitations are localized on a neighboring qubit, $L=1$. Fig.~\ref{fig:3}(a) presents this set of eigenvalues and its degree of localization of a $10\times10$ qubit lattice with $\varphi=3\pi/4$, where one can notice a cluster of correlated pairs highlighted by the dashed red circle. Naturally, the presence of non-radiative decay increases the collective decay rate for the entire set of states. Its effect on the subradiant states and on the polaritons pairs can be seen in Supplementary Materials. 

One additional way to characterize these polariton pairs is to investigate their entanglement. In particular, for bipartite systems, a powerful concept to measure the degree of entanglement between two quantum states is the von Neumann entanglement entropy, which can be defined as
\begin{eqnarray}
S &=&\frac{-\sum_{\nu} |\lambda_{\nu}|^2 \textrm{ ln} |\lambda_{\nu}|^2}{\sum_{\nu} |\lambda_{\nu}|^2}, \label{eq:entropy}
\end{eqnarray}
where $\lambda_{\nu}$ is the Schmidt coefficients obtained via the bipartite wave function rewritten using the
Schmidt decomposition as $\psi_{ij,nm}=\sum_{\nu}\lambda_{\nu}\psi_{i,j}\psi_{n,m}$.
It can be seen in Fig.~\ref{fig:3}(b) that the bound states correspond to the smallest entanglement entropy reflecting the fact that these states are closest to the pure two-particle states. It can be seen in Fig.~\ref{fig:3}(c), that for highly correlated pairs when the position of the first excitation is fixed (for example by the measurement), the second excitation is localized in the vicinity. Conversely, the subradiant states present a delocalized pattern as shown in Fig.~\ref{fig:3}(d). Noteworthy, the two-polariton pair class is not the most subradiant state of the system. At the same time, it is known that bound polariton pair lifetime depends crucially on the parameter $\varphi$~\cite{Poddubny2020}. It is therefore subject to further studies to check if there exists a \textit{magic} value of $\varphi$ for which the bound state becomes subradiant.

To conclude, we have shown that two-photon bound polariton states exist in two-dimensional WQED systems, and obtained their energy dispersion and spatial profile. These states result from the interplay between infinitely strong on-site repulsion of qubit excitations and strongly non-parabolic dispersion of the polariton modes, which contain regions corresponding to negative effective masses. The bound states exist in finite two-dimensional structures of modest size, and can be probed experimentally via the scattering measurements~\cite{gong2021quantum}. The recently demonstrated two-dimensional WQED systems based on superconducting qubits appear an ideal platform for the emulation of the lattice quantum walks and quantum Levy flights~\cite{deng2016quantum}. The character of the multiphoton bound states would influence the multi-photon transport  in these structures and thus plays a crucial role in their possible applications for quantum technologies.

\begin{acknowledgments}
The main results of the paper were obtained with the support of Russian Science Foundation (project 20-12-00224). The entanglement entropy has been calculated with the support of Russian Foundation of Basic research grant 20-02-00084. IAS acknowledges support from Icelandic Research Fund (project ”Hybrid polaritonics”). IVI acknowledges the support of "Basis" Foundation (project 21-1-2-61-1).
\end{acknowledgments}


\providecommand{\noopsort}[1]{}\providecommand{\singleletter}[1]{#1}%

\setcounter{figure}{0}
\setcounter{table}{0}
\setcounter{equation}{0}
\renewcommand{\theequation}{S\arabic{equation}}
\renewcommand\thefigure{S\arabic{figure}}
\section{Supplementary Materials}
\subsection{Derivation of the two-polariton Schrödinger equation}
\label{ap:A}

In this section, we provide the detailed derivation of the Schrödinger equation in the limit of two-particle excitation for the two-dimensional waveguide lattice, whose Hamiltonian is written as $\mathcal{H}_{\textrm{eff}}^{\textrm{2D}} = \mathcal{H}_{\textrm{eff}}^{\textrm{1D}}\otimes I+I\otimes\mathcal{H}_{\textrm{eff}}^{\textrm{1D}}$, 
where the 1D Hamiltonian is shown in Eq.(\ref{eq:H_1D}). Hence, the two-particle Schrödinger equation reads
\begin{eqnarray}
2\varepsilon\Psi & = & \left\{ \left[\left(H_{\textrm{1D}}\otimes I\right)+\left(I\otimes H_{\textrm{1D}}\right)\right]\otimes I^{\otimes2}\right\} \Psi\nonumber \\
 & + & \left\{I^{\otimes2}\otimes\left[\left(H_{\textrm{1D}}\otimes I\right)+\left(I\otimes H_{\textrm{1D}}\right)\right]\right\} \Psi, \label{eq:SE_2D}
\end{eqnarray}
in which the wave function
\begin{eqnarray}
\Psi & = & \sum_{i,j=1}^{N}\sum_{m,n=1}^{N}\psi_{ij,mn}b_{i,j}^{\dagger}b_{m,n}^{\dagger}\left|0\right\rangle
\end{eqnarray}
describes the two-particle state with the corresponding amplitude $\psi_{ij,mn}$ for the excitation pair respectively labeled by the indices $i,j$ and $m,n$, in which the indices $i,m$ correspond to the $x$-coordinates positions and $j,n$ to the $y$-coordinates positions. Substituting Eq.~(\ref{eq:H_1D}) into the Schrödinger equation (\ref{eq:SE_2D}), we obtain
\begin{eqnarray}
2\varepsilon\psi_{ij,mn} &=&  H_{il}\psi_{lj,mn}+H_{jl}\psi_{il,mn} +H_{ml}\psi_{ij,ln} \nonumber\\
&+&  H_{nl}\psi_{ij,ml}+\chi\delta_{im}\delta_{jn}\psi_{ij,mn}, \label{eq:SE_2D2}
\end{eqnarray}
where $H_{mn}=\omega_{0}\delta_{mn}-i\Gamma_{0}e^{i\varphi|m-n|}$ and we assumed a summation over the dummy index $l$. The energy $\omega_{0}$ is subtracted henceforward to shorten the notation as it contributes just as a Lamb-shift in the
eigenmodes. Within the hard-core boson limit ($\chi\rightarrow\infty$), $\psi_{ij,ij}\equiv0$ and $\chi\psi_{ij,ij}$ is a constant. In order to properly suppress $\chi$, we treat $\chi\psi_{ij,ij}$ as a perturbation by assuming $m=i$ and $n=j$ in Eq.~(\ref{eq:SE_2D2}) as follows
\begin{eqnarray}
\chi\psi_{ij,ij} &=& -H_{il}\psi_{lj,ij}-H_{il}\psi_{ij,lj}-H_{jl}\psi_{il,ij}-H_{jl}\psi_{ij,il}\nonumber\\
&=&-2H_{il}\psi_{ij,lj}-2H_{jl}\psi_{ij,il},\label{eq:const}
\end{eqnarray}
where we have used $\psi_{ij,mn}\equiv\psi_{mn,ij}$ as we are dealing with symmetric bosonic excitations. Therefore, the linear eigenvalue problem for the two-polariton excitation reads
\begin{align}
2\varepsilon\psi_{ij,mn}=&H_{il}\psi_{lj,mn}+H_{jl}\psi_{il,mn}+H_{ml}\psi_{ij,ln} \nonumber\\
+H_{nl}\psi_{ij,ml}-&2\delta_{jn}H_{il}\psi_{lj,in}-2\delta_{im}H_{jl}\psi_{il,mj}, \label{eq:TP_2D2}
\end{align}
which correspond to Eq.~(\ref{eq:TP_2D}) of the main text.
%
%

\subsection{Dispersion relation}
\label{ap:B}
\begin{figure}[t]
	\centerline{\includegraphics[width=3.4in,keepaspectratio]{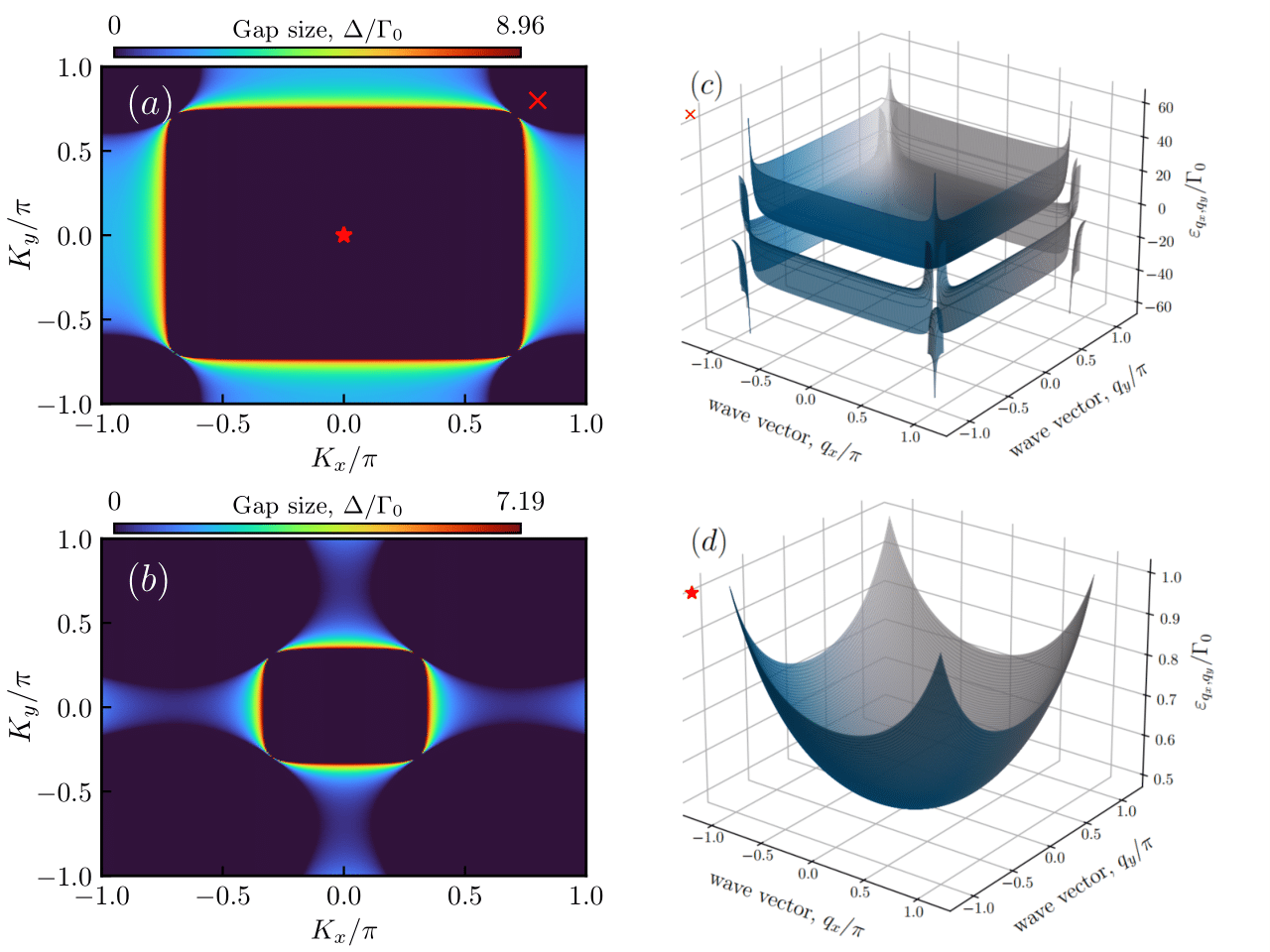}}
	\caption{(a), (b) The size of the energy gap in the polariton dispersion as a function of the center of mass wave vectors $\boldsymbol{K}$ for $\varphi=0.85\pi$ and $\varphi=0.65\pi$, respectively. (c), (d) polariton dispersion with $\varphi=0.85\pi$ respectively obtained for ${K_{x}=K_{y}=0.8\pi}$ (highlighted by a X) and ${K_{x}=K_{y}=0}$ (highlighted by a star).}\label{fig:s1}
\end{figure}

In this section, we present the scattering states energy in the domains where the energy gap is absent. Its expression, straightforwardly obtained by performing the 2D cosine Fourier transformation (guarantee symmetric bosonic excitations) in Eq.~(\ref{eq:ES_k}), is repeated below for convenience
\begin{eqnarray}
2\varepsilon_{q_{x},q_{y}}&=&\Gamma_{0}(\frac{\sin\varphi}{\cos k_{1,x}-\cos\varphi}+\frac{\sin\varphi}{\cos k_{2,x}-\cos\varphi}) \nonumber\\&+&\Gamma_{0}(\frac{\sin\varphi}{\cos k_{1,y}-\cos\varphi}+\frac{\sin\varphi}{\cos k_{2,y}-\cos\varphi})\label{eq:DR} 
\end{eqnarray}
 with polaritons wave vectors $k_{1(2),x(y)}=(q_{x(y)}\pm K_{x(y)})/2$. The dependence on the center of mass wave vectors $\boldsymbol{K}$ and phase $\varphi$ creates essentially three distinct dispersion shapes, one with a finite gap (shown in Fig.2(a) of the main text) and other two gapless. In this sense, Fig.~\ref{fig:s1} present these dispersions and points out in which domain they lie. Fig.~\ref{fig:s1}(a) and Fig.~\ref{fig:s1}(b) present the gap size $\Delta$ in the dispersion relation as a function of $\boldsymbol{K}$ for $\varphi=0.85\pi$ and $\varphi=0.65\pi$, respectively. Fig.~\ref{fig:s1}(c), obtained for ${K_{x}=K_{y}=0.8\pi}$, illustrates the shape of dispersion relation that are found for values of $\boldsymbol{K}$ that corresponds to the corners of both Fig.~\ref{fig:s1}(a) and Fig.~\ref{fig:s1}(b).

The profile of the dispersion relation found at the central regions are illustrated by Fig.~\ref{fig:s1}(d) that is obtained for ${K_{x}=K_{y}=0}$.\\ \\

%
%

%
\subsection{Real-space profile of the bound state}
Here we provide the evaluation of the integral in~\ref{eq:WF} for the specific case $K_x=\pi,K_y=0, \varphi=3\pi/4$. The integral of $q_x$ can be taken analytically yielding
\begin{align}
    \Phi_{d_x,d_y}=\pi\Gamma_0 (-1)^{d_x+1} \int_{-\pi}^{\pi} dq_y \tan^{|d_x|}\left(\frac{\xi(q_y)}{2}\right)\frac{\cos q_y d_y}{\sin (2\xi(q_y))},
\end{align}
for $d_x\neq 0$ and 
\begin{align}
    \Phi_{d_x,d_y}=-\pi\Gamma_0  \int_{-\pi}^{\pi} dq_y \tan\left(\frac{\xi(q_y)}{2}\right)\frac{\cos q_y d_y}{\cos (\xi(q_y))},
\end{align}
for $d_x=0$,where
\begin{align}
\xi(q_y)=\arcsin\left[\varepsilon_b-\frac{1}{1+\sqrt{2}\cos(q_y/2)}\right]. 
\end{align}
Naturally, $\Phi_{0,0}=0$, and it can be seen that the wave function decays as $e^{-\alpha|d_x|}$.

\subsection{Non-radiative decay}
\label{ap:C}

\begin{figure}[t]
	\centerline{\includegraphics[width=3.4in,keepaspectratio]{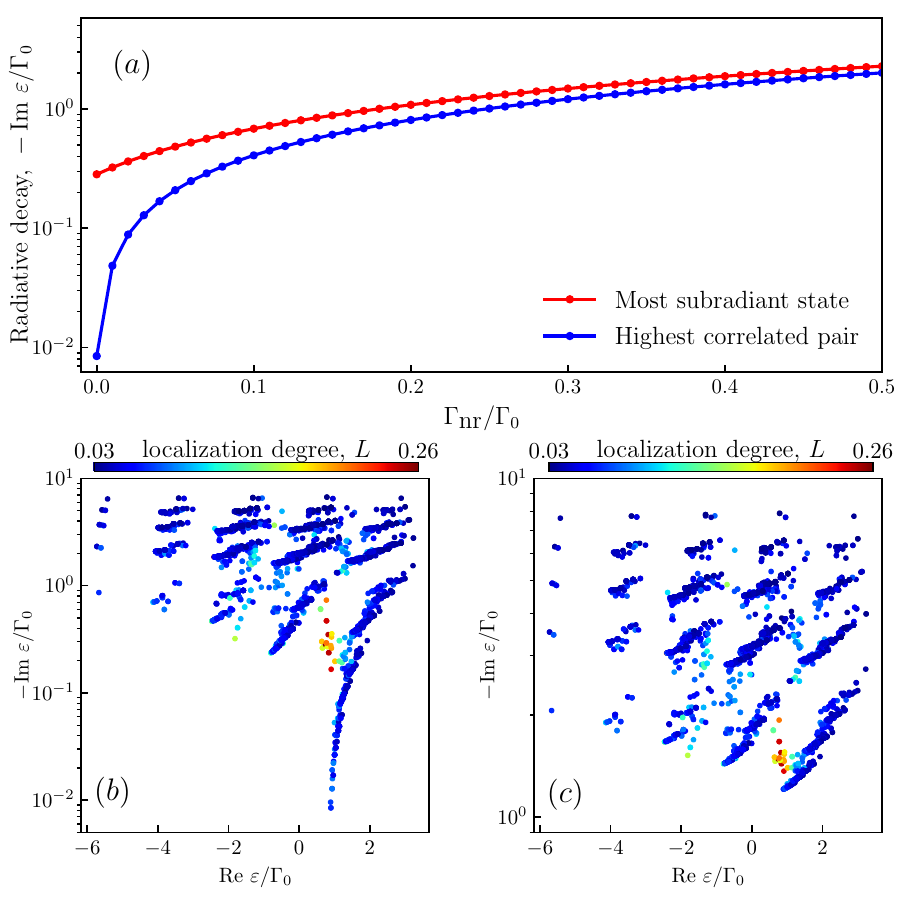}}
	\caption{(a) Emission decay rate of the highest correlated pair and the most subradiant states as a function of coupling ratio $\Gamma_{\textrm{nr}}/\Gamma_{0}$. (b) , (c) The two-excitation eigenvalues of the finite structure system composed of $N^{2}=64$ qubits characterized by the localization degree $L$ of its wave functions for $\Gamma_{\textrm{nr}}=0$ and $\Gamma_{\textrm{nr}}=0.3\Gamma_{0}$, respectively. The polariton phase is set at $\varphi=3\pi/4$.}\label{fig:s2}
\end{figure}

One of the most challenging tasks in waveguide QED systems is to fabricate experimental set-ups where photons remain guided and thus prevented to be emitted to the free space. Although that is an experimental reality for superconductor qubits~\cite{Mirhosseini2019} and, at a certain level, also for artificial atoms coupled to nanofiber waveguides~\cite{Corzo2019}, that is not the case for most waveguide QED systems~\cite{sheremet2021waveguide}. In our model, this effect is taken into account by the non-radiative decay rate $\Gamma_{\textrm{nr}}$ of each individual qubit as shown by Hamiltonian (2) of the main text, namely
\begin{eqnarray}
\mathcal{H}_{\textrm{eff}}^{\textrm{1D}} & = & \sum_{m,n=1}^{N}[(\omega_{0}-i\Gamma_{\textrm{nr}})\delta_{mn}-i\Gamma_{0}e^{i\varphi|m-n|}]b_{m}^{\dagger}b_{n} \nonumber\\
& + & \frac{\chi}{2} \sum_{m=1}^{N} b_{m}^{\dagger} b_{m}^{\dagger}b_{m}b_{m},\label{eq:H_1Ds} 
\end{eqnarray}
where both $\Gamma_{0}$  and $\Gamma_{\textrm{nr}}$ correspond to photon decay to the waveguide and to the free space, respectively. As these two competing channels of energy loss,  are described by the imaginary terms they naturally vanish when the periodic lattice condition is met in infinite systems. To properly investigate this competition in the two-polariton state formation as well as in the corresponding collective emission rate, $-{\textrm{Im}(\varepsilon)}$, in a finite set-up, we present the evolution of the most localized polariton pair and the most subradiant state emission rate as a function of coupling efficiency for lattice with $N^2=64$ atoms in Fig.~\ref{fig:s2}. 

Notably, one can notice in Fig~\ref{fig:s2}(a) that the collective emission decay of both the highest correlated pair and the most subsradiant state are equally increased by the coupling ratio $\Gamma_{\textrm{nr}}/\Gamma_{0}$. Fig~\ref{fig:s2}(b) display the complete set of eigenvalues considered an idealized situation of maximum efficiency ($\Gamma_{\textrm{nr}}=0$), while Fig~\ref{fig:s2}(c) show that the collective emission is highly increased for $\Gamma_{\textrm{nr}}=0.3\Gamma_{0}$. Moreover, the non-radiate losses does not affect the localization degree as can be seen by  Fig~\ref{fig:s2}(b) and Fig~\ref{fig:s2}(c).

\end{document}